\documentclass[11pt,a4paper]{article}
\usepackage{amsmath}
\usepackage{amssymb}
\usepackage{amsthm}
\usepackage{a4wide,fullpage}
\usepackage{fancybox}
\usepackage{stmaryrd}
\usepackage{tabularx}
\usepackage{makeidx}
\usepackage[colorlinks=true,linkcolor=blue,citecolor=blue,urlcolor=blue,pdfauthor={Philippe Gambette, Stéphane Vialette},pdftitle={On restrictions of balanced 2-interval graphs}]{hyperref}
\usepackage{enumerate}
\usepackage[T1]{fontenc}
\usepackage[dvips]{graphicx}
\makeindex

\newtheorem{property}{Property}
\newtheorem{lemma}{Lemma}
\newtheorem{theorem}{Theorem}

  \renewenvironment{thebibliography}[1]{%
    \begin{oldthebibliography}{#1}%
      \setlength{\parskip}{0ex}%
      \setlength{\itemsep}{0ex}%
  }%
  {%
    \end{oldthebibliography}%
  }

\title{On restrictions of balanced 2-interval graphs}
\author{%
  Philippe Gambette  \\
  LIAFA, UMR CNRS 7089, Universit\'e Paris 7, France \\
  D\'epartement Informatique, ENS Cachan, France \\
  \texttt{gambette@liafa.jussieu.fr}
  \and
  St\'ephane Vialette \\
  LRI, UMR CNRS 8623, Universit\'e Paris-Sud 11, France \\
  \texttt{vialette@lri.fr}
 }
\date{}

\begin{document}
\maketitle

\begin{abstract}
The class of 2-interval graphs has been introduced for modelling
scheduling and allocation problems, 
and more recently for specific bioinformatic problems. 
Some of those applications  
imply restrictions on the 2-interval graphs, and justify the introduction
of a hierarchy of subclasses of 2-interval graphs that generalize line
graphs: balanced 
2-interval graphs, unit 2-interval graphs, and ($x$,$x$)-interval
graphs. We provide 
instances that show that all the inclusions are strict. We extend the
NP-completeness proof of recognizing 2-interval graphs to the
recognition of balanced 2-interval 
graphs. Finally we give hints on the complexity of unit 2-interval
graphs recognition, 
by studying relationships with other graph classes: proper
circular-arc, quasi-line graphs, $K_{1,5}$-free graphs, \ldots 

\end{abstract}

\section{2-interval graphs and restrictions}\label{intro}

The interval number of a graph, and the classes of $k$-interval graphs
have been introduced as a generalization of the class of interval
graphs by McGuigan~\cite{MCGuigan1977} in the context
of scheduling and allocation problems.
Recently, bioinformatics problems have renewed interest in
the class of 2-interval graphs (each vertex is associated to a pair of
disjoint intervals and edges denote intersection between two such
pairs).
Indeed, a pair of intervals can
model two associated tasks in scheduling~\cite{BHNSS2006}, but also two
similar segments of DNA in the context of DNA comparison~\cite{JMT1992},
or two complementary segments of RNA for RNA secondary structure
prediction and comparison~\cite{Vialette:2004}.

\begin{figure}[!htb]
\centering
\begin{tabular}{ccc}
\includegraphics[scale=0.45]{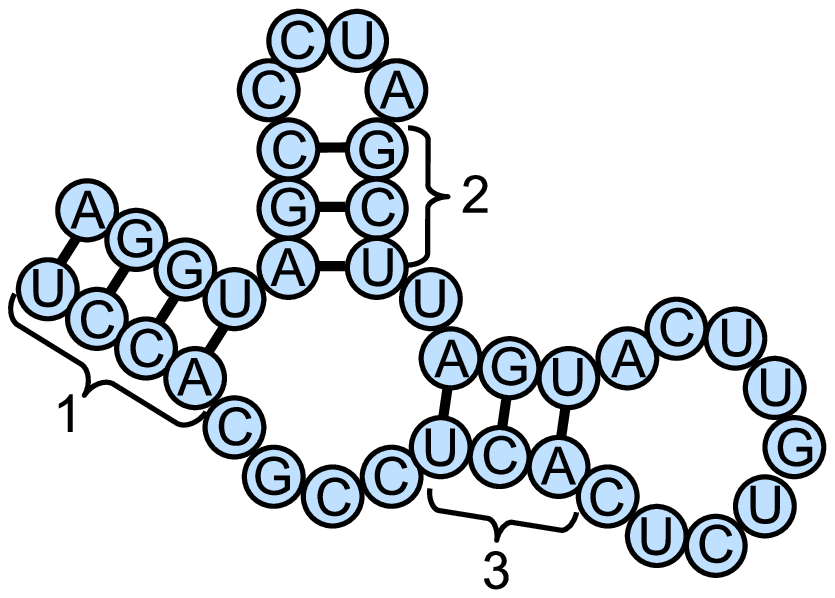} &
\includegraphics[scale=0.45]{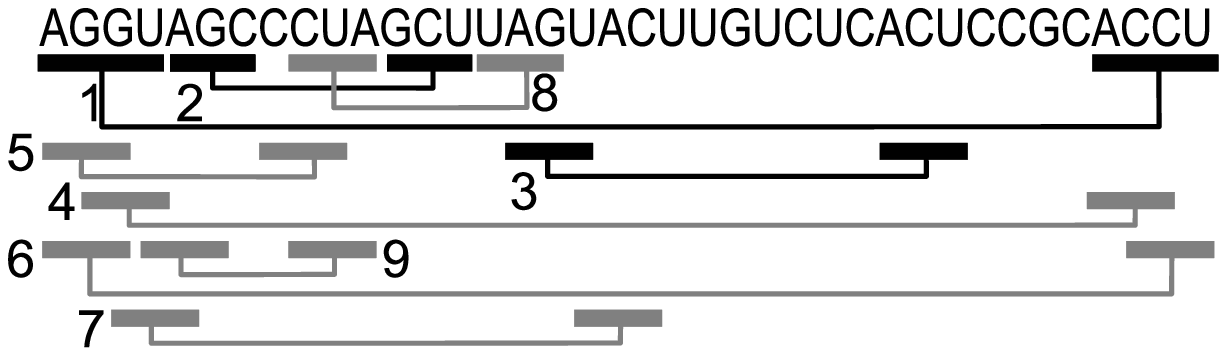} &
\includegraphics[scale=0.45]{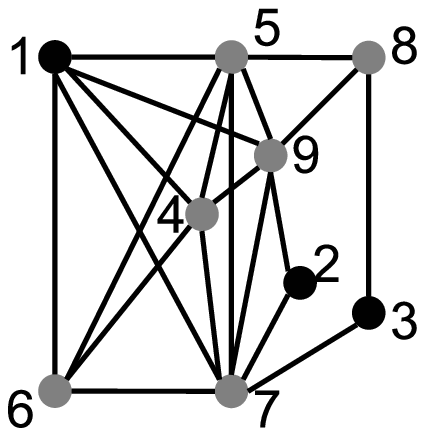}\\
(a) & (b) & (c)
\end{tabular}
\caption{Helices in a RNA secondary structure (a) can be modeled
as a set of balanced 2-intervals among all 2-intervals corresponding
to complementary and inverted pairs of letter sequences (b),
or as an independent subset in the balanced associated 2-interval graph (c).}
\label{FigArn}
\end{figure}
RNA (ribonucleic acid) are polymers of nucleotides linked in a chain
through phosphodiester bonds. 
Unlike DNA, RNAs are usually single stranded, but many RNA molecules
have \emph{secondary structure} in which intramolecular loops are formed by
complementary base pairing. 
RNA secondary structure is generally divided into helices (contiguous
base pairs), and various kinds of loops (unpaired nucleotides
surrounded by \emph{helices}). 
The structural stability and function of non-coding RNA (ncRNA) genes
are largely determined by the formation of stable secondary structures 
through complementary bases, and hence
ncRNA genes across different species are most similar in the pattern
of nucleotide complementarity rather than in the genomic sequence.
This motivates the use of 2-intervals for modelling RNA secondary
structures: each helix of the structure is modeled by a 2-interval. 
Moreover, the fact that these 2-intervals are usually required to be
disjoint in the structure naturally suggests the use of 2-interval
graphs.  
Furthermore, aiming at better modelling RNA secondary structures, it
was suggested in~\cite{CHLV2005} to focus on 
\emph{balanced 2-interval sets} (each 2-interval is composed of two
equal length intervals) and their associated intersection graphs 
referred as \emph{balanced 2-interval graphs}.
Indeed, helices in RNA secondary structures are most of the time
composed of equal length contiguous base pairs parts.
To the best of our knowledge, nothing is known on the class of
balanced 2-interval graphs. 

Sharper restrictions have also been introduced in scheduling,
where it is possible to consider tasks which all have the same
duration, that is 2-interval whose intervals have the same
length~\cite{BHNSS2006,Karlsson2005}. This motivates the 
study of the classes of unit 2-interval graphs, and $(x,x)$-interval
graphs. In this paper, we consider these subclasses of interval
graphs, and in particular we address the problem of recognizing them.

A graph $G=(V,E)$ of order $n$ is a 2-interval graph if it is the
intersection graph of a set of $n$ unions of two disjoint intervals on
the real line, that is each vertex corresponds to a
union of two disjoint intervals $I^k = I^k_l \cup I^k_r$,
$k \in \llbracket 1,n \rrbracket$ ($l$ for ``\emph{left}'' and 
$r$ for ``\emph{right}''), and there is an edge between $I^j$ and
$I^k$ iff $I^j \cap I^k \neq \emptyset$. 
Note that for the sake of simplicity we use the same letter
to denote a vertex and its corresponding 2-interval.
A set of 2-intervals corresponding to a graph $G$ is called a realization of $G$.
The set of all intervals, $\bigcup_{k=1}^{n} \{I^k_l,I^k_r\}$, is called
the ground set of $G$ (or the ground set of $\{I^1,\ldots,I^n\}$).

The class of 2-interval graphs is a generalization of interval graphs,
and also contains all circular-arc graphs (intersection graphs of arcs of a circle),
outerplanar graphs (have a planar embedding with all vertices around
one of the faces~\cite{KostochkaWest1999}),
cubic graphs (maximum degree 3~\cite{GriggsWest1980}),
and line graphs (intersection graphs of edges of a graph).

Unfortunately, most classical graph combinatorial problems turn out to
be NP-complete for 2-interval graphs:
recognition~\cite{WestShmoys1984}, maximum independent
set~\cite{BNR1996,Vialette2001}, 
coloration~\cite{Vialette2001}, \ldots 
Surprisingly enough, the complexity of the maximum clique problem for
2-interval graphs is still open (although it has been recently proven
to be NP-complete for 3-interval graphs~\cite{BHLR2007}).

\begin{figure}[!htb]
\centering
\includegraphics[scale=0.45]{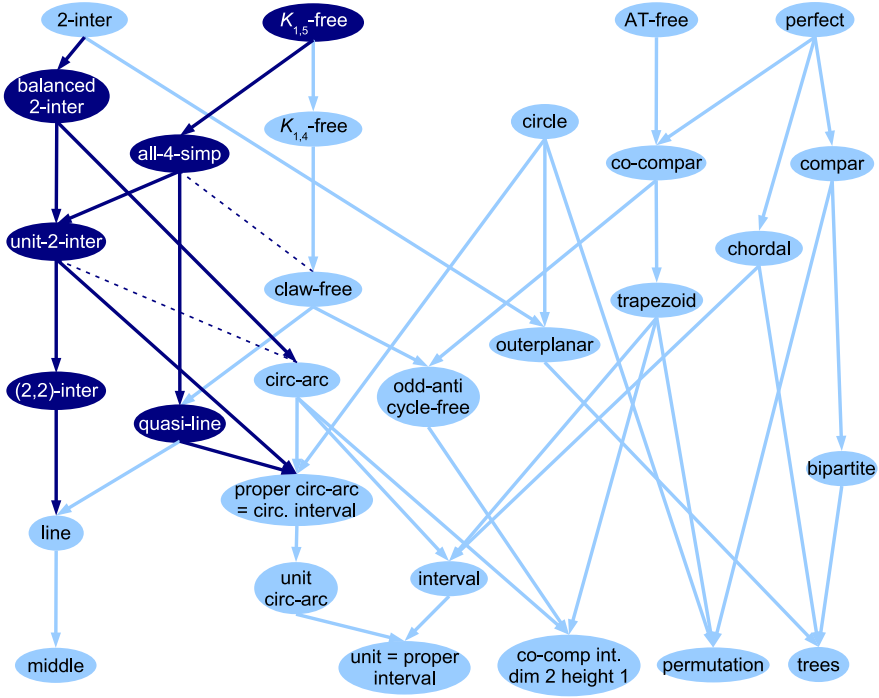}
\caption{Graph classes related to 2-interval graphs
and its restrictions. A class pointing towards another strictly contains it,
and the dashed lines mean that there is no inclusion relationship
between the two. Dark classes correspond to classes not yet present
in the ISGCI Database~\cite{ISGCI}.}
\label{FigGraphClasses}
\end{figure}

For practical application, restricted 2-interval graphs are needed. 
A 2-interval graph is said to be \emph{balanced} if it has a
2-interval realization in which each 2-interval is composed of two
intervals of the same length~\cite{CHLV2005},
\emph{unit} if it has a 2-interval realization in which all intervals
of the ground set have length 1~\cite{BFV2004}, 
and is called a \emph{$(x,x)$-interval graph} if it has a 2-interval
realization in which all intervals of the ground set 
are open, have integer endpoints, and length
$x$~\cite{BHNSS2006,Karlsson2005}. 
In the following sections, we will study those restrictions of
2-interval graphs, and their position in the hierarchy of graph classes
illustrated in Figure~\ref{FigGraphClasses}.

Note that all $(x,x)$-interval graphs are unit 2-interval graphs,
and that all unit 2-interval graphs are balanced 2-interval graphs.
We can also notice that $(1,1)$-interval graphs are exactly line graphs:
each interval of length 1 of the ground set can be considered as the vertex
of a root graph and each 2-interval as an edge in the root graph.
This implies for example that the coloration problem is also
NP-complete for $(2,2)$-interval graphs and wider classes of graphs.
It is also known that the complexity of the maximum independent set problem
is NP-complete on $(2,2)$-interval graphs~\cite{BNR1996}.
Recognition of $(1,2)$-union graphs, a related class (restriction
of \emph{multitrack interval graphs}),
was also recently proven NP-complete~\cite{HalldorsonKarlsson2006}.

\section{Useful gadgets for 2-interval graphs and restrictions}

\begin{figure}[!htb]
\centering
\begin{tabular}{ccc}
\includegraphics[scale=0.55]{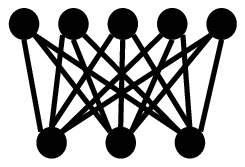}
& \includegraphics[scale=0.55]{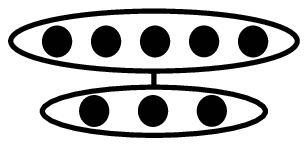}
& \includegraphics[scale=0.55]{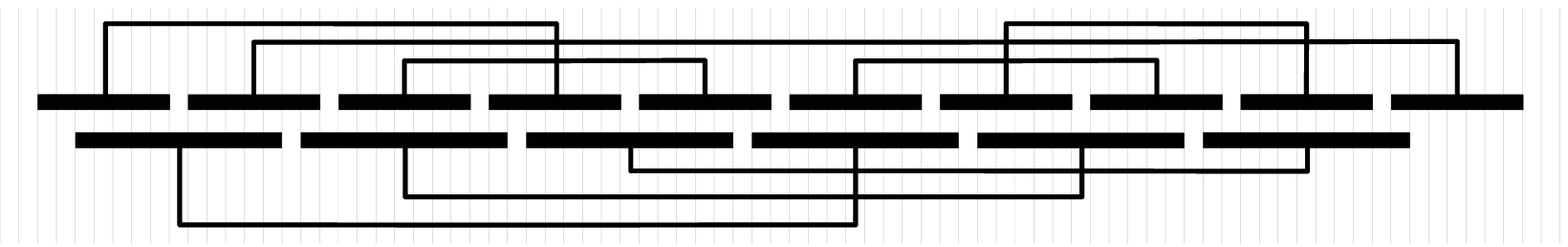}
\\
(a) & (b) & (c)
\end{tabular}
\caption{The complete bipartite graph $K_{5,3}$ (a,b) has a
balanced 2-interval realization (c): vertices of $S_5$
are associated to balanced 2-intervals of length $7$, 
and vertices of $S_3$ are associated to balanced 2-intervals of length
$11$. 
Any realization of this graph is contiguous, \emph{i.e.},
the union of all 2-intervals is an interval.} 
\label{FigK53eq}
\end{figure}

For proving hardness of recognizing 2-interval graphs, West and Shmoys
considered in~\cite{WestShmoys1984} the complete bipartite graph
$K_{5,3}$ as a useful 2-interval gadget.
Indeed, all realizations of this graph are contiguous, that is, for
any realization, the union of all intervals in its ground set is an
interval. 
Thus, by putting edges between 
some vertices of a $K_{5,3}$ and another vertex $v$, we can force
one interval of the 2-interval $v$ (or just one extremity of
this interval) to be blocked inside the realization of $K_{5,3}$.
It is not difficult to see that $K_{5,3}$ has a balanced 2-interval
realization, for example the one in Figure~\ref{FigK53eq}.

However, $K_{5,3}$ is not a unit 2-interval graph.
Indeed, each 2-interval $I = I_l \cup I_r$ corresponding to a degree 5
vertex intersect $5$ disjoint 2-intervals, and hence one of $I_l$ or
$I_r$ intersect at least $3$ intervals, which is impossible for unit
intervals. 
Therefore, we introduce the new gadget $K_{4,4}-e$
which is a $(2,2)$-interval graph with only contiguous realizations.

\begin{figure}[!htb]
\centering
\begin{tabular}{ccc}
\includegraphics[scale=0.55]{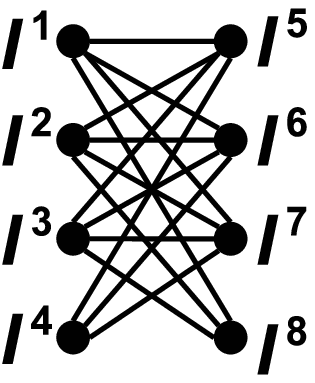}
& \includegraphics[scale=0.55]{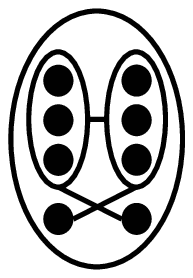}
& \includegraphics[scale=0.55]{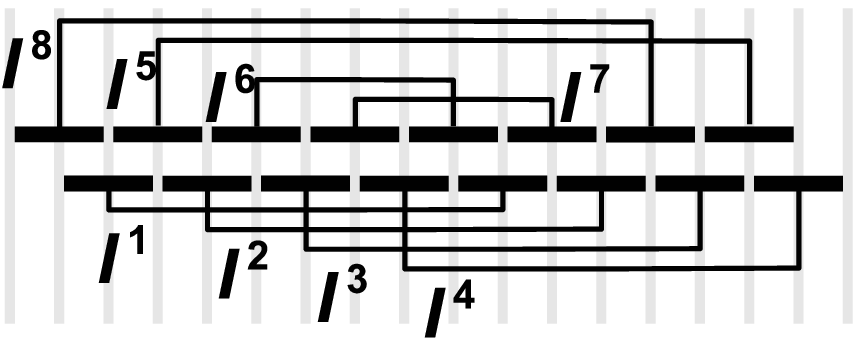}
\\
(a) & (b) & (c)
\end{tabular}
\caption{The graph $K_{4,4}-e$ (a), a nicer representation (b),
and a 2-interval realization with open intervals of length 2 (c).}
\label{FigK4-e}
\end{figure}

\begin{property}
Any 2-interval realization of $K_{4,4}-e$ is contiguous.
\end{property}
\begin{proof}
Write $G=(V,E)$ the graph $K_{4,4}-e$.
To study all possible realizations of $G$, let us study
all possible realizations of $G[V-I^8]$.

As 2-intervals $I^1$, $I^2$, $I^3$ and $I^4$ are disjoints,
their ground set $\mathcal{I}_\textrm{fixed}=\{[l_i,r_i],$
$1\leq i\leq 8,$ $r_i < l_{i+1}\}$
is a set of eight disjoint intervals.
The ground set $\mathcal{I}_\textrm{mobile}$ of $I^5$, $I^6$ and $I^7$
is a set of six disjoint intervals.
Let $x_i$ be the number of intervals of $\mathcal{I}_\textrm{mobile}$ 
intersecting $i \leq 8$ intervals of $\mathcal{I}_\textrm{fixed}$.
We have directly:
\begin{equation}\label{eqReal1}
x_0 + x_1 + x_2 + x_3 + x_4 + x_5 + x_6 + x_7 + x_8 = |\mathcal{I}_\textrm{mobile}| = 6.
\end{equation}

As there are 12 edges in $G[V \backslash \{v_8\}]$ which is bipartite, we also have:
\begin{equation}\label{eqReal2}
x_1 + 2 x_2 + 3 x_3 + 4 x_4 + 5 x_5 + 6 x_6 + 7 x_7 + 8 x_8  \geq 12.
\end{equation}

%
%
Finally, to build a realization of $G$ from
a realization of $G[V \backslash \{v_8\}]$ , one must place $I^8$ so as to intersect
three disjoint intervals of $\mathcal{I}_\textrm{fixed}$.
Thus one of the intervals of $I^8$ intersects at least two intervals
$]l_k,r_k[$ 
and $]l_l,r_l[$ ($k<l$) of $\mathcal{I}_\textrm{fixed}$. So there is
``a hole between those two intervals'', for example $[r_k,l_{k+1}]$, 
which is included in one of the intervals of $I^8$.
So we notice that $I^8$ has to fill one of the seven
holes of $\mathcal{I}_\textrm{fixed}$.
Thus, the intervals of $\mathcal{I}_\textrm{mobile}$ can not fill
more than six holes, and the observation that an interval intersecting
$i$ consecutive intervals (for $i \geq 1$) fills $i-1$ holes, we get:
\begin{equation}\label{eqReal4}
x_2 + 2 x_3 + 3 x_4 + 4 x_5 + 5 x_6 + 6 x_7 + 7 x_8 \leq 6.
\end{equation}

Equations~\ref{eqReal1}, \ref{eqReal2} 
and \ref{eqReal4} are necessary for any valid realization of 
$G[V \backslash \{v_8\}]$ which gives a valid realization of $G$.

Let's suppose by contradiction that the union of all intervals of the ground
set of $G$ is not an interval. Then there is a hole, that is an interval
included in the covering interval of $\{I^1,\ldots,I^8\}$, which
intersect no $I^i$. We proceed like for equation~\ref{eqReal4},
with the constraint that another hole cannot be filled by the
intervals of $\mathcal{I}_\textrm{mobile}$, so we get instead:
\begin{equation}\label{eqReal4bis}
x_2 + 2 x_3 + 3 x_4 + 4 x_5 + 5 x_6 + 6 x_7 + 7 x_8 \leq 5.
\end{equation}

By adding \ref{eqReal1} and \ref{eqReal4bis}, and subtracting \ref{eqReal2}, 
we get $x_0 \leq -1$ : impossible! So we have proved that
the union of all intervals of the ground
set of any realization of $G$ is indeed an interval.
\end{proof}

\section{Balanced 2-interval graphs}

We show in this section that the class of balanced 2-interval graphs is
strictly included in the class of 2-interval graphs, and strictly contains
circular-arc graphs.
Moreover, we prove that recognizing balanced 2-interval graphs is as
hard as recognizing (general) 2-interval graphs.

\begin{property}\label{balanced}
The class of balanced 2-interval graphs is strictly included
in the class of 2-interval graphs.
\end{property}
\begin{proof}
We build a 2-interval graph that has no balanced 2-interval realization.
Let's consider a chain of gadgets $K_{5,3}$ (introduced in previous section)
to which we add three vertices $I^1$, $I^2$, and $I^3$ as illustrated
in Figure~\ref{FigContrexemple2}.
\begin{figure}[!htb]
\centering
\begin{tabular}{c}
(a)\\
\includegraphics[scale=0.55]{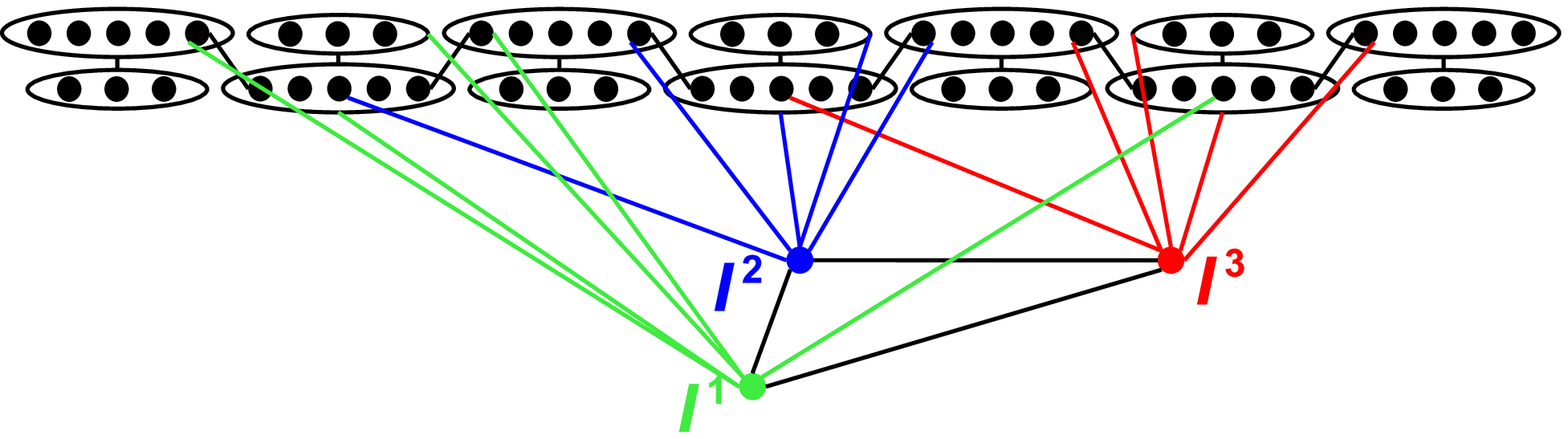}\\
(b)\\
\includegraphics[scale=0.55]{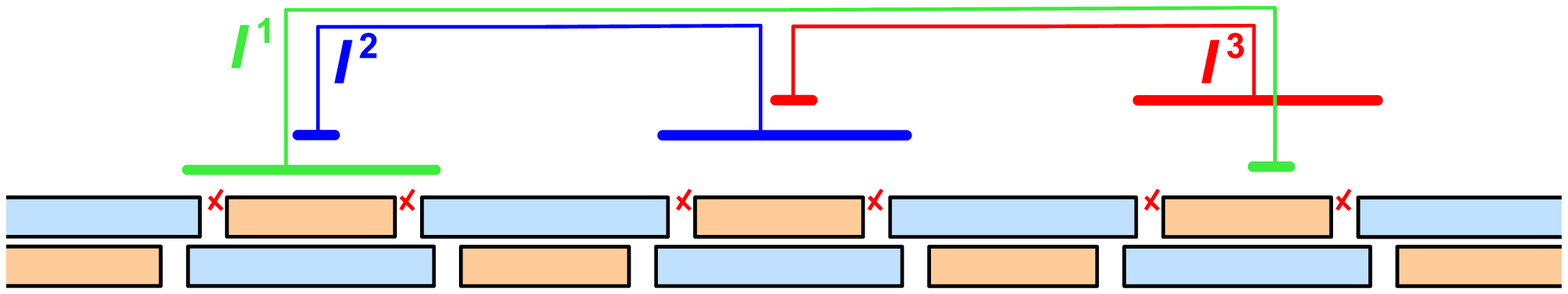}
\end{tabular}
\caption{An example of unbalanced 2-interval graph~(a)~:
any realization groups intervals of the seven $K_{5,3}$ in
a block, and the chain of seven blocks creates six ``holes'' between them,
which make it impossible to balance the lengths of
the three 2-intervals $I^1$, $I^2$, and $I^3$.
}
\label{FigContrexemple2}
\end{figure}

In any realization, the presence of holes showed by crosses in the Figure
gives the following inequalities for any realization:
$l({I_l}^2) < l({I_l}^1)$, $l({I_l}^3) < l({I_r}^2)$,
and $l({I_r}^1) < l({I_r}^3)$ (or if the realization of the chain of
$K_{5,3}$ appears in the symmetrical order:
$l({I_l}^1) < l({I_l}^3)$, $l({I_r}^3) < l({I_l}^2)$,
and $l({I_r}^2) < l({I_r}^1)$).
If this realization was balanced, then we would have
$l({I_l}^1) = l({I_r}^1) < l({I_r}^3) = l({I_l}^3) < l({I_r}^2) = l({I_l}^2)$
(or for the symmetrical case:
$l({I_r}^1) = l({I_l}^1) < l({I_l}^3) = l({I_r}^3) < l({I_l}^2) = l({I_r}^2)$) :
impossible! So this graph has no balanced 2-interval realization although it has
a 2-interval generalization.
\end{proof}

\begin{theorem}\label{balancedNPcomplete}
Recognizing balanced 2-interval graphs is an NP-complete problem.
\end{theorem}
\begin{proof}
We just adapt the proof of West and Shmoys~\cite{WestShmoys1984,GyarfasWest1995}.
The problem of determining if there is a Hamiltonian cycle in a 3-regular
triangle-free graph is proven NP-complete, by reduction from the more general
problem without the no triangle restriction. So we reduce the problem of Hamiltonian cycle
in a 3-regular triangle-free graph to balanced 2-interval recognition.

Let $G=(V,E)$ be a 3-regular triangle-free graph. We build
a graph $G'$ which has a 2-interval realization (a special one, very
specific, called $H$-representation and which we prove to be balanced)
iff $G$ has a Hamiltonian cycle.
\begin{figure}[!htb]
\centering
\begin{tabular}{c}
\includegraphics[scale=0.55]{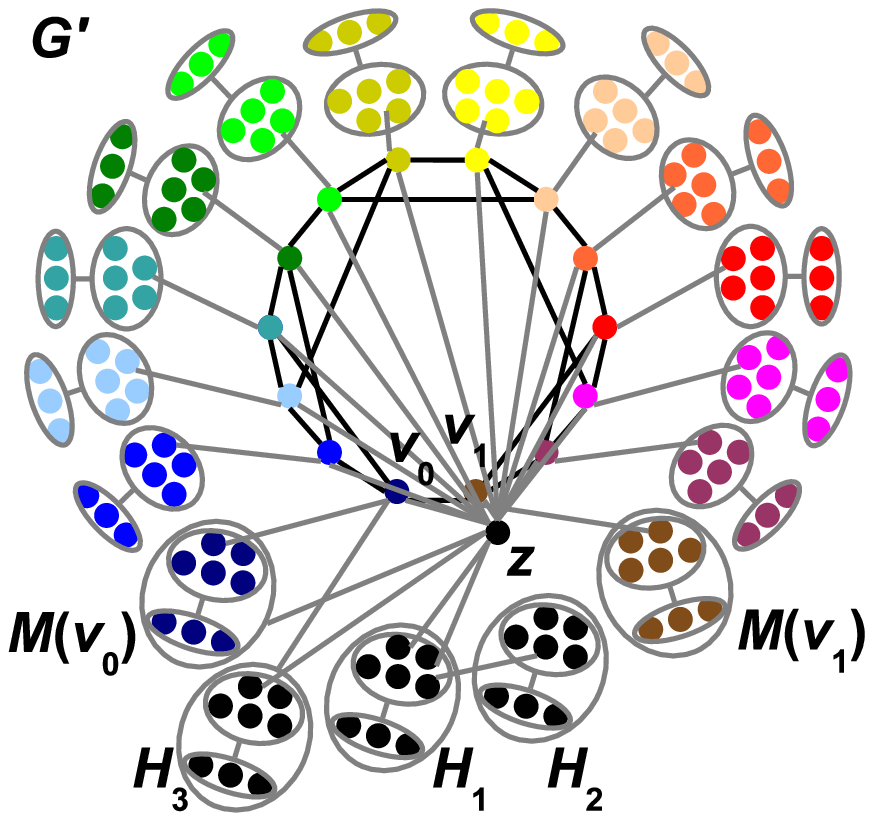}\\
\includegraphics[scale=0.55]{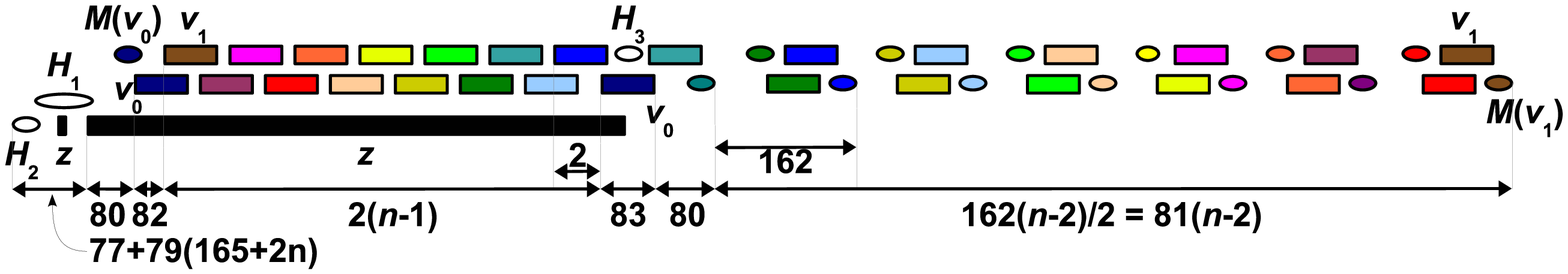}
\end{tabular}
\caption{There is a balanced 2-interval of $G'$ (which has been
dilated in the drawing to remain readable) iff there
is an $H$-representation (that is a realization
where the left intervals of all 2-intervals are
grouped together in a contiguous  block) for its induced subgraph $G$
iff there is a Hamiltonian cycle in $G$.
}
\label{FigReductionWestShmoysEq}
\end{figure}

The construction of $G'$, illustrated in Figure~\ref{FigReductionWestShmoysEq}(a)
is almost identical to the one by West and Shmoys,
so we just prove that $G'$ has a balanced realization,
shown in Figure~\ref{FigReductionWestShmoysEq} (b), by computing
lengths for each interval to ensure it.
All $K_{5,3}$ have a balanced realization as shown in section~\ref{intro}
of total length 79, in particular $H_3$.
We can thus affect length 83 to the intervals of $v_0$.
The intervals of the other $v_i$ can have length 3, 
and their $M(v_i)$ length 79, so through the computation illustrated in
Figure~\ref{FigReductionWestShmoysEq}, intervals of $z$
can have length $80+82+2(n-1)+3$, that is $163+2n$.
We dilate $H_1$ until a hole between two consecutive intervals of its $S_3$
can contain an interval of $z$, that is until the hole has length $165+2n$ :
so after this dilating, $H_1$ has length $79(165+2n)$.
Finally if $G$ has a Hamiltonian cycle, then we have found a balanced
2-interval realization of $G$ of total length $13,273+241n$.
\end{proof}

It is known that circular-arc graphs are 2-interval graphs,
they are also balanced 2-interval.
\begin{property}\label{circularArcBalanced}
The class of circular-arc graphs is strictly included in the class
of balanced 2-interval graphs.
\end{property}
\begin{proof}
The transformation is simple: if we have a circular-arc
representation of a graph $G=(V,E)$, then we choose some point $P$
of the circle. We partition $V$ in $V_1 \cup V_2$, where
$P$ intersects all the arcs corresponding to vertices of $V_1$ and
none of the arcs of the vertices of $V_2$. Then we cut the circle
at point $P$ to map it to a line segment: every arc of $V_2$ becomes
an interval, and every arc of $V_1$ becomes a 2-interval.
To obtain a balanced realization we just cut in half the intervals of $V_2$
to obtain two intervals of equal length for each. And for each 2-interval
$[g(I_l),d(I_l)]\cup[g(I_r),d(I_r)]$ of $V_1$, as both intervals are
located on one of the extremities of the realization, we can increase the
length of the shortest so that it reaches the length of the longest without
changing intersections with the other intervals.
The inclusion is strict because $K_{2,3}$ is a balanced 2-interval graph (as
a subgraph of $K_{5,3}$ for example) but is not a circular-arc graph (we
can find two $C_4$ in $K_{2,3}$, and only one can be realized with
a circular-arc representation).
\end{proof}

\section{Unit 2-interval and (x,x)-interval graphs}

\begin{property}\label{InclusionXX}
Let $x \in \mathbb{N}, x \geq 2$. The class of $(x,x)$-interval graphs
is strictly included in the class of $(x+1,x+1)$-interval graphs.
\end{property}
\begin{proof}
We first prove that an interval graph with a representation where all
intervals have length $k$ (and integer open bounds) has a
representation where all intervals have length $k+1$. 

We use the following algorithm. Let $S$ be initialized as
the set of all intervals of length $k$, and let $T$ be initially the
empty set. 
As long as $S$ is not empty,
let $I=[a,b]$ be the left-most interval of $S$, remove from $S$
each interval $[\alpha,\beta]$ such that $\alpha < b$ (including $I$),
add $[\alpha,\beta +1]$ to $T$, and translate by +1 all the remaining intervals
in $S$.
When $S$ is empty, the intersection graph of $T$, where all intervals
have length 
$k+1$ is the same as the intersection graph for the original $S$.

We also build for each $x \geq 2$ a $(x+1,x+1)$-interval graph which is not
a $(x,x)$-interval graph. We consider the bipartite graph
$K_{2x}$ and a perfect matching 
$\{(v_i,v'_i), i\in \llbracket 1,x \rrbracket\}$.
We call $K'_x$ the graph obtained from $K_{2x}$ with the following transformations,
illustrated in Figure~\ref{FigxxInter}(a):
remove edges $(v_i,v'_i)$ of the perfect matching,
add four graphs $K_{4,4}-e$ called $X_1$, $X_2$, $X_3$, $X_4$
(for each $X_i$, we call $v_l^i$ and $v_r^i$ the vertices of degree 3),
link $v_r^2$ and $v_l^3$, link all $v_i$ to $v_r^1$ and $v_l^4$,
link all $v'_i$ to $v_l^2$ and $v_r^3$, and finally add a vertex $a$ (resp. $b$)
linked to all $v_i$, $v'_i$, and to two adjacent vertices
of $X_1$ (resp. $X_4$) of degree 4.
We illustrate in Figure~\ref{FigxxInter}(b) that $K'_x$
has a realization with intervals of length $x+1$.
We can prove by induction on $x$ that $K'_x$ has no realization
with intervals of length $x$: it is rather technical,
so we just give the idea. Any realization of $K'_x$ forces
the block $X_2$ to share an extremity with the block $X_3$,
so each 2-interval $v'_i$ has one interval intersecting the other extremity
of $X_2$, and the other intersecting the other extremity of $X_3$.
Then constraints on the position of vertices $v_i$ force
their intervals to appear as two ``stairways'' as shown in Figure~\ref{FigxxInter}(b).
So $v_r^1$ must contain $x$ extremities of intervals which have to be different,
so it must have length $x+1$.
\begin{figure}[!htb]
\centering
(a)\\
\includegraphics[scale=0.55]{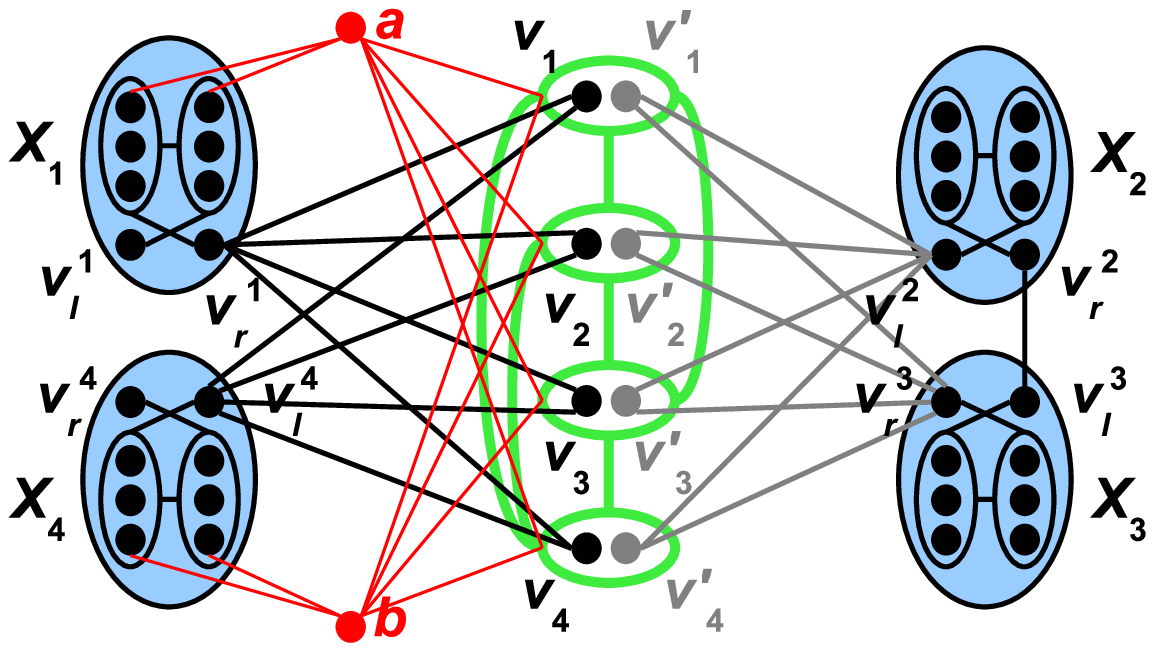}\\
~\\
(b)\\
\includegraphics[scale=0.55]{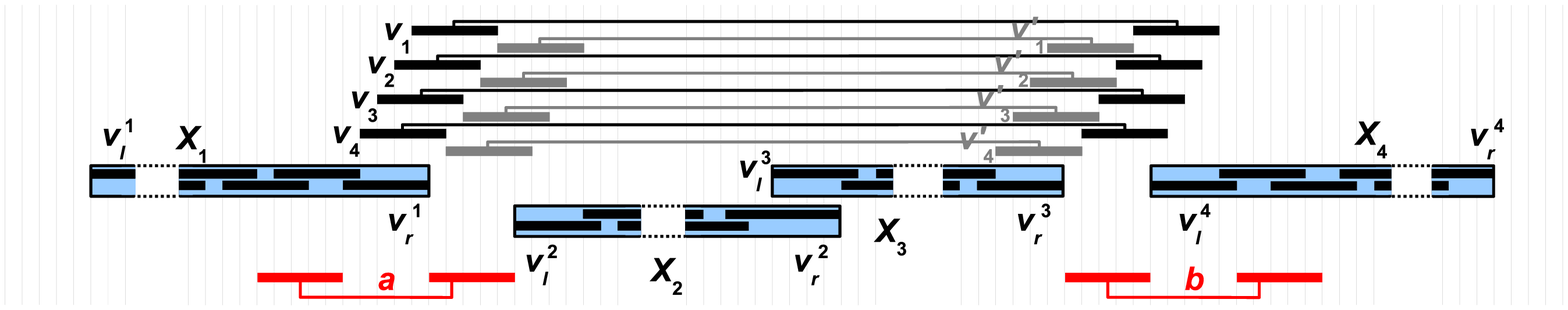}
\caption{The graph $K'_4$ (a) is (5,5)-interval
but not (4,4)-interval.}
\label{FigxxInter}
\end{figure}
\end{proof}

The complexity of recognizing unit 2-interval graphs and $(x,x)$-interval
graphs remains open, however the following shows a relationship
between those complexities.
\begin{lemma}
$\{$unit 2-interval graphs$\} =
\bigcup\limits_{x \in \mathbb{N}^*} \{(x,x)$-interval graphs$\}$.
\end{lemma}
\begin{proof}
The $\supset$ part is trivial.
To prove $\subset$, let $G=(V,E)$ be a unit 2-interval graph.
Then it has a realization
with $|V|=n$ 2-intervals, that is $2n$ intervals of the ground set.
So we consider the interval graph of the ground set, which is
a unit interval graph.
There is a linear time algorithm based on breadth-first search
to compute a realization of such a graph where interval
endpoints are rational, with denominator $2n$~\cite{CKNOS1995}. So by dilating
by a factor $2n$ such a realization, we obtain a realization
of $G$ where intervals of the ground set have length $2n$.
\end{proof}

\begin{theorem}
If recognizing $(x,x)$-interval graphs is polynomial for any integer $x$
then recognizing unit 2-interval graphs is polynomial.
\end{theorem}

\section{Investigating the complexity of unit 2-interval graphs}

In this section we show that all proper circular-arc graphs
(circular-arc graphs such that no arc is included in another
in the representation) are unit 2-interval graphs, and we study a class of graphs
which generalizes quasi-line graphs and contains unit 2-interval graphs.

Recall that, according to Property~\ref{circularArcBalanced}, 
circular-arc graphs are balanced 2-interval graphs. 
However, circular-arc graphs are not necessarily unit 2-interval
graphs.

\begin{property}\label{unitquasi}
The class of proper circular-arc graphs is strictly included in the class
of unit 2-interval graphs.
\end{property}
\begin{proof}
As in the proof of Property~\ref{circularArcBalanced}, we choose
a point $P$ on the circle of the representation of a proper
circular-arc graph $G$, 
and maps the cut circle into a line segment. We extend the outer extremities
of intervals that have been cut so that no interval contains another.
Thus we obtain a set of 2-intervals for arcs containing $P$,
and a set $I$ of intervals for arcs not containing $P$. For each
interval of $I$, 
we add a new interval disjoint of any other to get a 2-interval.
If we consider the intersection graph of the ground set of such a
representation, 
it is a proper interval graph. So it is also a unit interval graph~\cite{Roberts1969}, which
provides a unit 2-interval representation of $G$.

To complete the proof, we notice that the domino (two cycles $C_4$
having an edge in common) is a unit 2-interval graph but not a
circular-arc graph. 
\end{proof}

Quasi-line graphs are those graphs whose vertices are bisimplicial,
\emph{i.e.}, the closed neighborhood of each vertex is the union of
two cliques. 
This graph class has been introduced as a generalization of line
graphs and a useful subclass of claw-free
graphs~\cite{BenRebea1981,FFR1997,ChudnovskySeymour2005,KingReed2007}.
Following the example of quasi-line graphs that generalize line
graphs, we introduce here a new class of graphs for generalizing unit
2-interval graphs. 
Let $k \in \mathbb{N}^*$. A graph $G=(V,E)$ is 
\emph{all-$k$-simplicial} if the neighborhood of each vertex $v \in V$
can be partitioned into at most $k$ cliques. 
The class of quasi-line graphs is thus exactly the class of
all-$2$-simplicial graphs. 
Notice that this definition is equivalent to the following:
in the complement graph of $G$, for each vertex $u$, the vertices that
are not in the neighborhood of $u$ are $k$-colorable.

\begin{property}\label{unitall4}
The class of unit 2-interval graphs is strictly included in the class
of all-4-simplicial graphs.
\end{property}
\begin{proof}
The inclusion is trivial.
What is left is to show that the inclusion is strict.
Consider the following graph which is all-$4$-simplicial but not unit 
2-interval: 
start with the cycle $C_4$, 
call its vertices $v_i$, $i \in \llbracket 1,4\rrbracket$,  
add four $K_{4,4}-e$ gadgets called $X_i$, and for each $i$
we connect the vertex $v_i$ to two connected vertices of degree 4 in
$X_i$. 
This graph is certainly all-$4$-simplicial.
But if we try to build a 2-interval realization of this graph, then
each of the 2-intervals $v_k$ has an interval trapped into
the block $X_k$. So each 2-interval $v_k$ has only one interval 
to realize the intersections with the other $v_i$: this
is impossible as we have to realize a $C_4$ which has no interval
representation. 
\end{proof}

\begin{property}\label{clawquasi}
The class of claw-free graphs is not included in the class
of all-4-simplicial graphs.
\end{property}
\begin{proof}
The Kneser Graph $KG(7,2)$ is triangle-free, but not
4-colorable~\cite{Lovasz1978}. 
We consider the graph obtained by adding an isolated vertex $v$
and then taking the complement graph, 
\emph{i.e.}, $\overline{KG(7,2) \uplus \{v\}}$.
It is claw-free as $KG(7,2)$ is triangle-free. And if it was
all-$4$-simplicial, 
then the neighborhood of $v$ in $\overline{KG(7,2) \uplus \{v\}}$, that
is $\overline{KG(7,2)}$, would be a union of at most four cliques, so
$KG(7,2)$ would be 4-colorable: impossible so this graph is claw-free but
not all-4-simplicial.
\end{proof}

\begin{property}
The class of all-$k$-simplicial graphs is strictly included in the
class of $K_{1,k+1}$-free graphs.
\end{property}
\begin{proof}
If a graph $G$ contains $K_{1,k+1}$, then it has a vertex
with $k+1$ independent neighbors, and hence $G$ is not
all-$k$-simplicial. 
The wheel $W_{2k+1}$ is a simple example of a graph which is $K_{1,k+1}$-free 
but in which the center can not have its neighborhood (a $C_{2k+1}$) 
partitioned into $k$ cliques or less.
\end{proof}

Unfortunately, all-$k$-simplicial graphs do not have a nice structure
which could help unit 2-interval graph recognition.  

\begin{theorem}
Recognizing all-$k$-simplicial graphs is NP-complete for $k \geq 3$.
\end{theorem}
\begin{proof}
We reduce from the \textsc{Graph $k$-colorability} problem, which is
known to be NP-complete for $k\geq 3$~\cite{Karp1972}.
Let $G=(V,E)$ be a graph, and let $G'$ be the complement
graph of $G$ to which we add a universal vertex $v$. 
We claim that $G$ is $k$-colorable iff
$G'$ is all-$k$-simplicial.

If $G$ is $k$-colorable, then the non-neighborhood of any vertex in
$G$ is $k$-colorable, so the neighborhood of any vertex in $\overline G$ is
a union of at most $k$ cliques. And the neighborhood of $v$ is also
a union of at most $k$ cliques, so $G'$ is all-$k$-simplicial.

Conversely, if $G'$ is all-$k$-simplicial, then in particular the
neighborhood of 
$v$ is a union of at most $k$ cliques. Let's partition it into $k$
vertex-disjoint cliques $X_1,\ldots, X_k$. Then, coloring $G$ such
that two vertices have the same color iff they are in the same $X_i$
leads to a valid $k$-coloring of $G$.
\end{proof}



\section{Conclusion}

Motivated by practical applications in scheduling and computational
biology, we focused in this paper on balanced 2-interval graphs and
unit 2-intervals graphs.
Also, we introduced two natural new classes: 
$(x,x)$-interval graphs and all-$k$-simplicial graphs.

We mention here some directions for future works.
First, the complexity of recognizing unit 2-interval graphs and
$(x,x)$-interval graphs remains open.
Second, the relationships between quasi-line graphs and subclasses of 
balanced 2-intervals graphs still have to be investigated.  
Last, since most problems remains NP-hard for balanced 2-interval
graphs, there is thus a natural interest in investigating the
complexity and approximation of classical optimization problems on
unit 2-interval graphs and $(x,x)$-interval graphs.

\section*{Acknowledgments}

We are grateful to Vincent Limouzy in particular for bringing to our attention
the class of quasi-line graphs, and Michael Rao and Michel Habib for useful discussions.

\small{
\bibliographystyle{alpha}
\bibliography{Bib2006}
}

\normalsize
\newpage
\section{Appendix}

We give the detailed proofs of Theorem~\ref{balancedNPcomplete} and
Property~\ref{InclusionXX}.
\begin{proof}[Proof of Theorem~\ref{balancedNPcomplete}]
Let $G=(V,E)$ be a 3-regular triangle-free graph. We build
a graph $G'$ which has a 2-interval realization (a special one, very
specific, and which we prove to be balanced) iff $G$ has a Hamiltonian cycle.

First we will detail how we build $G'$ starting from the graph $G$,
and adding some vertices, in particular $K_{5,3}$ gadgets.
The idea is that the edges of $G$ will partition into a Hamiltonian cycle
and a perfect matching iff all 2-intervals of the realization of $G'$
can have their left interval realizing the Hamiltonian cycle,
and their right interval realizing the perfect matching.
A realization with such a placement of the intervals
is called an ``H-representation'' of $G$.

We proceed as illustrated in Figure~\ref{FigReductionWestShmoysEq}.
We choose some vertex of $G$ that we call $v_0$ (which will be the ``origin''
of the Hamiltonian cycle), and the other are called $v_1, \ldots, v_n$.
For each vertex $v_i$ of $G$ we link it to a vertex of the $S_5$ of a $K_{5,3}$
called $M(v_i)$ (which will block one of the four extremities of the 2-interval $v_i$).
We link all vertices to a new vertex $z$, which is linked to no $M(v)$
except $M(v_0)$ (thus the interval of each $v_i$ intersecting $M(v_i)$,
for $i\neq 0$, won't intersect $z$). We add three $K_{5,3}$, $H_1$, $H_2$
and $H_3$ : two vertices of the $S_5$ of $H_1$ are linked to $z$,
a third one is linked to one vertex of the $S_5$ of $H_2$, one vertex
of the $S_5$ of $H_3$ is linked to $z$, and all vertices of $H_3$ to $v_0$.

To explain this construction in detail, we study the realization of $G'$,
if we suppose it is a (balanced) 2-interval graph, and we
prove that it leads us to find a Hamiltonian cycle in $G$.

As the realization of $H_1$ and $H_2$ are two contiguous blocks of intervals
then one of their extremities must intersect.
As $z$ is linked to two disjoint vertices of $H_1$, both intervals
of $z$ are used to realize those intersections. But one interval
of $z$ that we call $z_r$, also has to intersect one vertex of $H_3$
which is not linked to $H_1$,
so $z_r$ intersects the second extremity of the block $H_1$
(the first extremity being occupied by the extremity of $H_2$).
And as $z_r$ intersects only one interval of $H_3$,
it must be the extremity of $H_3$.
The other interval of $z$ is contained in the block $H_1$,
thus can't intersect $M(v_0)$ neither all
the vertices $v_i$, so all those 2-intervals intersect $z_r$.
And as none of them intersect $H_3$ except $v_0$, all of them
except $v_0$ have an interval contained in $z_r$,
that we call $v_{i,g}$.
The other interval of each $v_i$ is linked to a $K_{5,3}$
so it has one extremity occupied by $K_{5,3}$, and the other one
is free.

Conversely, if $G$ has a Hamiltonian cycle, then it is possible to
find a $H$-representation, such that all the constraints induced
by the edges of $G'$ are respected, as illustrated with the realization
in Figure~\ref{FigReductionWestShmoysEq}. We have already proved that this
realization can be balanced.
\end{proof}

\begin{proof}[Proof of Property~\ref{InclusionXX}]

In the following, as we only considering the interval of $v_l^i$ or
$v_r^i$ located at one extremity of the block $X_i$, and not the one
inside, we will use $v_l^i$ and $v_r^i$ to denote those extremity
intervals. 
For each vertex $v_i$, we call $v_{i,l}$ its left interval and
$v_{i,r}$ its right interval. We do the same for $v'_i$, and
call $l(I)$ the left extremity of any interval $I$.

We prove by induction that the graph $K'_x$ is $(x+1,x+1)$-interval but
not $(x,x)$-interval, and that for any unit 2-interval realization,
there exists an order $\sigma \in \mathcal{S}_x$ such that :
\begin{itemize}
\item either $l(v_{\sigma(x),l})< \ldots < l(v_{\sigma(1),l}) < l(v'_{\sigma(x),l})
< \ldots < l(v'_{\sigma(1),l})$ and $l(v'_{\sigma(x),r})< \ldots < l(v'_{\sigma(1),r})
< l(v_{\sigma(x),r}) < \ldots < l(v_{\sigma(1),r})$,
\item or the symmetric case: $l(v_{\sigma(1),l}) < \ldots < l(v_{\sigma(x),l})
< l(v'_{\sigma(1),l}) < \ldots < l(v'_{\sigma(x),l})$ and
$l(v'_{\sigma(1),r}) < \ldots < l(v'_{\sigma(x),r})
< l(v_{\sigma(1),r}) < \ldots < l(v_{\sigma(x),r})$.
\end{itemize}
Those two equalities correspond in fact to the ``two stairways structure'' which
appears in Figure~\ref{FigxxInter}.
~\\
\textbf{Base case} :
we study all possible unit 2-interval realizations of $K'_2$
to prove that one of the expected inequalities is always true.
We also prove that $K'_2$ has no (2,2)-interval realization.

First recall that realizations of $X_i$ subgraphs can only be blocks
of contiguous intervals. The edge between $v_r^2$ and $v_l^3$ forces
the two blocks of $X_2$ and $X_3$ to be contiguous, with intervals
$v_l^2$ and $v_r^3$ at their extremities.
Each 2-interval $v'_i$ must intersect both $v_l^2$ and $v_r^3$,
so one of its intervals intersects $v_l^2$ and the other intersects $v_r^3$.
Thus, one same interval of $v'_i$ can not intersect both $a$ and $b$
which are disjoint, so $a$ intersects one interval of $v'_i$ (say the one
intersecting $v_l^2$, the other case being treated symmetrically)
and $b$ intersects the other one (so, the one intersecting $v_r^3$).
Each $v_i$ has to intersect both $a$ and $b$, so it has to intersect
$a$ with its first interval and $b$ with the second.
But 2-interval $v_i$ must also intersect $v^1_r$ and $v^4_l$ which are
both disjoint and disjoint to $a$ and $b$. So one interval of each $v_i$
must intersect $v^1_r$ and the other one must intersect $v^4_l$.

So we have shown that any unit 2-interval realization of $K'_2$
has the following aspect (or the symmetric) : the extremity of the block $X_1$
intersecting all $v_i$ which intersect $a$ (or $b$) which intersects
all $v'_i$, which intersect the extremity $X_2$ (or $X_3$) which intersects
the extremity of $X_3$ (or $X_2$), which intersects all $v'_i$, which intersect $b$
(or $a$), which intersects all $v_i$, which intersect the extremity of $X_4$.

Now we suppose, by contradiction, that there exists a (2,2)-interval realization
of $K'_2$. $v_r^1$ is an interval of length 2, but one of its two parts
of length one has to intersect an element of $X_1$. The other has to intersect both
$v_1$ and $v_2$. As neither $v_1$ nor $v_2$ can intersect other intervals
of $X_1$, then the first interval of $v_1$ and $v_2$ is the same interval.
By proceeding the same way on $X_4$ and $v_l^4$, we obtain that the second
interval of $v_1$ and $v_2$ is the same interval, so $v_1$ and $v_2$
should correspond to the same 2-interval: it contradicts with the
fact that vertices $v_1$ and $v_2$ have a different neighborhood.
So $K'_2$ has no (2,2)-interval realization.

To obtain the expected inequalities, we have to analyze the
possible positions of all $v_i$ and $v'_i$.
We only treat the first two inequalities as the second case is symmetric.

Suppose that $l(v_{2,l})<l(v_{1,l})$. As $v_1$ and $v'_1$
are non adjacent, then interval $v_{1,l}$ is strictly on the left
of $v'_{1,l}$, so $v_{2,l}$ is strictly on the left
of $v'_{1,l}$. Thus those two intervals do not intersect.
But $v_2$ and $v'_1$ are adjacent, so $v_2$ and $v'_1$
must have intersecting right intervals. But then we have
$l(v'_{2,r})<l(v'_{1,r})<l(v_{2,r})<l(v_{1,r})$, and the right
intervals of $v'_2$ and $v_1$ can not intersect.
We deduce their left intervals intersect,
so $l(v_{2,l})<l(v_{1,l})<l(v'_{2,l})<l(v'_{1,l})$.

If we suppose that $l(v_{1,l})<l(v_{2,l})$, we get as well
that $l(v'_{1,r})<l(v'_{2,r})<l(v_{1,r})<l(v_{2,r})$
and $l(v_{1,l})<l(v_{2,l})<l(v'_{1,l})<l(v'_{2,l})$. 
So for any unit 2-interval realization of $K'_2$ there exists an order
$\sigma= 1 2$ or $\sigma = 2 1$ such that:
\begin{itemize}
\item either $l(v_{\sigma(2),l})<l(v_{\sigma(1),l})<l(v'_{\sigma(2),l})<l(v'_{\sigma(1),l})$
and $l(v'_{\sigma(2),r})<l(v'_{\sigma(1),r})<l(v_{\sigma(2),r})<l(v_{\sigma(1),r})$,
\item or the symmetric inequalities.
\end{itemize}
~\\

\textbf{Recursion:}
suppose that for some $x$, $K'_{x-1}$ is not $(x-1,x-1)$-interval but
is $(x,x)$-interval, and that any $(x,x)$-interval realization
verifies one of the expected inequalities.

Graph $K'_{x-1}$ is an induce subgraph of $K'_{x}=(V,E)$ :
$K'_{x-1} = K'_{x}[V \setminus \{v_x,v'_x\}]$.
So by the induction hypothesis, there exists an order
$\sigma \in \mathcal{S}_{x-1}$ such that for any
unit 2-interval realization of $K'_{x}$ :
\begin{itemize}
\item either $l(v_{\sigma(x-1),l})< \ldots < l(v_{\sigma(1),l}) < l(v'_{\sigma(x-1),l})
< \ldots < l(v'_{\sigma(1),l})$ and $l(v'_{\sigma(x-1),r})< \ldots < l(v'_{\sigma(1),r})
< l(v_{\sigma(x-1),r}) < \ldots < l(v_{\sigma(1),r})$,
\item or the symmetric case: $l(v_{\sigma(1),l}) < \ldots < l(v_{\sigma(x-1),l})
< l(v'_{\sigma(1),l}) < \ldots < l(v'_{\sigma(x-1),l})$ and
$l(v'_{\sigma(1),r}) < \ldots < l(v'_{\sigma(x-1),r})
< l(v_{\sigma(1),r}) < \ldots < l(v_{\sigma(x-1),r})$.
\end{itemize}

The position of $v_x$ and $v'_x$ remains to be determined.
We treat only the case where the first two inequalities are true,
as the second case is symmetric.

As $v_x$ and $v_r^1$ are adjacent,
and $v'_{\sigma(x-1)}$ and $v_r^1$ are not,
then $l(v_r^1) < l(v_{x,l}) < l(v'_{\sigma(x-1),l})$. So
we define $j$ the following way: $v_{\sigma(j),l}$
is the leftmost interval 
such that $l(v_{x,l}) \leq l(v_{\sigma(j),l})$.
if there is none, we say $j=0$.
Then we call $\sigma' \in \mathcal{S}_{x}$ the permutation defined by:
\begin{displaymath}
\left\{
\begin{array}{l}
\sigma'(i)=\sigma(i)\textrm{ if }i<j,\\
\sigma'(j+1)=x,\\
\sigma'(i)=\sigma(i-1)\textrm{ if }i>j.
\end{array}\right.
\end{displaymath}

Then we directly get inequalities:
\begin{itemize}
\item $l(v_r^1) < l(v_{\sigma'(x),l}) < \ldots
< l(v_{\sigma'(j+1),l}) \leq l(v_{x,l}) < l(v_{\sigma'(j-1),l})
< \ldots < l(v_{\sigma'(1),l}) < l(v'_{\sigma'(x),l}) < \ldots
< l(v'_{\sigma'(j+1),l}) < l(v'_{\sigma'(j-1),l})
< \ldots < l(v'_{\sigma'(1),l})$
\item $l(v'_{\sigma'(x),r}) < \ldots
< l(v'_{\sigma'(j+1),r}) < l(v'_{\sigma'(j-1),r})
< \ldots < l(v'_{\sigma'(1),r})
< l(v_{\sigma'(x),r}) < \ldots
< l(v_{\sigma'(j+1),r}) < l(v_{\sigma'(j-1),r})
< \ldots < l(v_{\sigma'(1),r})$
\end{itemize}

We obtain the expected inequalities by reasoning the same way
as in the end of the base case.

So in particular we have $l(v_{\sigma(x),l})< \ldots < l(v_{\sigma(1),l})$
and $v_r^1$ must intersect all those $v_i$ for $i \in \llbracket 1,x \rrbracket$,
but also an interval of $X_1$ which intersects none of the $v_i$.
So it must have length $x+1$, thus $K'_x$ is not a $(x,x)$-interval graph

\textbf{Conclusion:} As the base case and the recursion has been proved,
expected properties of the graph $K'_x$ are true for any $x \geq 2$.
\end{proof}
\end{document}